\documentclass[conference]{IEEEtran}
\IEEEoverridecommandlockouts
\usepackage{cite}
\usepackage{amsmath,amssymb,amsfonts}
\usepackage{algorithm}
\usepackage{algorithmic}
\usepackage{graphicx}
\usepackage{textcomp}

\usepackage{colortbl} 
\usepackage[table]{xcolor} 
\usepackage{subcaption} 
\usepackage{url}

\def\BibTeX{{\rm B\kern-.05em{\sc i\kern-.025em b}\kern-.08em
    T\kern-.1667em\lower.7ex\hbox{E}\kern-.125emX}}
\begin{document}

\title{Autonomous Agricultural Monitoring with Aerial Drones and RF Energy-Harvesting Sensor Tags}

\author{
\IEEEauthorblockN{Paul S. Kudyba and Haijian Sun 
}
\IEEEauthorblockA{School of Electrical and Computer Engineering,
University of Georgia, 
Athens, GA, USA \\
paul.kudyba@uga.edu, hsun@uga.edu
}
}
\maketitle

\begin{abstract} 
In precision agriculture and plant science, there is an increasing demand for wireless sensors that are easy to deploy, maintain, and monitor. This paper investigates a novel approach that leverages recent advances in extremely low-power wireless communication and sensing, as well as the rapidly increasing availability of unmanned aerial vehicle (UAV) platforms. By mounting a specialized wireless payload on a UAV, battery-less sensor tags can harvest wireless beacon signals emitted from the drone, dramatically reducing the cost per sensor. These tags can measure environmental information such as temperature and humidity, then encrypt and transmit the data in the range of several meters. An experimental implementation was constructed at AERPAW, an NSF-funded wireless aerial drone research platform. While ground-based tests confirmed reliable sensor operation and data collection, airborne trials encountered wireless interference that impeded successfully detecting tag data. Despite these challenges, our results suggest further refinements could improve reliability and advance precision agriculture and agrarian research.
\end{abstract}


\section{Introduction} 
Driven by the growing demands of precision agriculture and plant research, the integration of secure, cost-effective, and battery-free wireless tags presents a promising enhancement to existing drone-assisted farming systems \cite{tsouros2019review}. Aerial drone platforms are already widely used in arable farming for tasks such as fertilizer application, weed management, and pest control. Additionally, environmental sensors are becoming increasingly prevalent in agricultural settings. However, conventional radio-based Internet of Things (IoT) sensors often fail to meet farmers' practical requirements. These limitations, which are inherently interconnected in agricultural applications, have hindered widespread adoption, making direct implementations of existing wireless technologies insufficient for meeting the evolving needs of modern farming.

The cost of current IoT devices for agricultural use limits sensor deployment to strategic locations within a field, which may be suitable for certain applications, such as monitoring soil health or weather conditions. However, these stationary sensors, along with their supporting infrastructure, impose ongoing maintenance costs that often exceed their initial material costs. Maintenance demands, including battery replacements, repositioning for optimal data collection, and retrieval to prevent damage from routine field operations like harvesting, place a significant burden on farmers' time and resources, which further limits their practicality and adoption \cite{9453805}. Current agricultural sensors are deployed within the coverage constraints of wireless standards such as Long Range (LoRa) \cite{ghazali2021systematic}, 802.11ah (Wi-Fi HaLow) \cite{alam2022ieee}, and private 5G gateways \cite{li2018uav}. To ensure connectivity, the necessary infrastructure needs to be invested to support these networks. However, they often face unexpected range limitations due to environmental factors that impact wireless propagation, as well as the ongoing challenge of maintaining battery-powered sensors. These limitations not only increase operational costs but also discourage widespread adoption, as farmers may find the technology unreliable or insufficient for large-scale deployment.

Emerging wireless technologies, coupled with the growing societal demand for data-driven solutions to enhance food security, present new opportunities for integrated IoT systems that align farmer and consumer trust with reliable agricultural data. In this paper, we explore the collection of temperature and identification data, both of which play critical roles in modern farming and supply chain management. Temperature data enables valuable inferences, such as assessing plant health and predicting product spoilage, while secure identification facilitates end-to-end traceability from farm to table, which can strengthen transparency and trust in the supply chain. 

    \begin{figure}[ht] 
        \centering
        \includegraphics[width=0.9\linewidth]{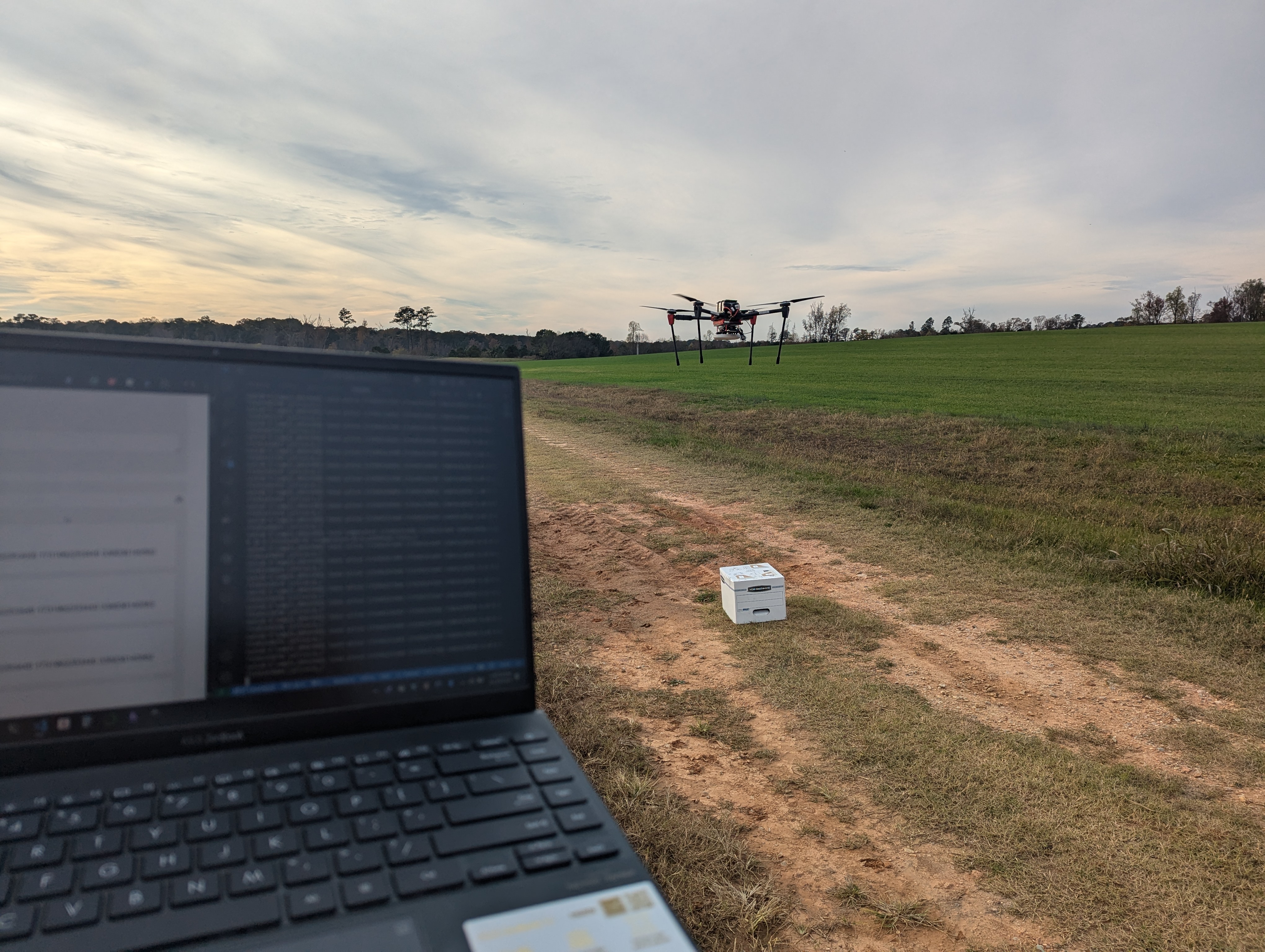} 
        \caption{A manual drone flight hovering above the target box with tags attached. On the left, a laptop was used to monitor data collection.} 
        \label{fig:man_flight}
    \end{figure}

This approach envisions automating data collection with sensor tags that flow with the product or object of study instead of requiring significant sensor material and maintenance costs. To obtain a sensor data point, the tag device is momentarily powered by an aerial drone. This is done by leveraging automated drone flights to deliver power and collect sensor data simultaneously \cite{8422277}. The data is encrypted to be secure at rest or used within broadcast wireless communication. Removing the need for battery energy storage in each sensor tag reduces costs.  Correspondingly, the communication range becomes constant as an aerial drone can rapidly move within an optimal receiving range for each sensor \cite{yangEnergyEfficiencyOptimization2019}. Because many agricultural processes that require monitoring evolve slower than most other IoT-sensed processes, this approach reduces extreme oversampling \cite{10757589}. This approach also avoids sensor misconfigurations or complex duplex communication schemes required to update polling intervals by using the drone as a surveyor. 
The drone automation schedule determines the sample rate required to establish a specific target inference, paving the way for greater automation in the future.

The rest of the paper is structured as follows. The wireless tag technology is described in Section \ref{sec:wirelessTag}. In Section \ref{sec:AERPAW}, an overview of the AERPAW platform is given. 
Section \ref{sec:method} details the specific approaches and methods used to construct the physical radio payload, setup a backend for data collection, and program an automated routine for the Small AERPAW Multirotor (SAM).
Experimental results are shown in Section \ref{sec:results} along with how to adapt to unforeseen circumstances. Finally, concluding remarks are given in Section \ref{sec:conclusion} and further directions follow in Section~\ref{sec:further}.

\subsection{Wireless Tags} \label{sec:wirelessTag} 
Passive wireless tags have seen numerous advances in their applications. Initially used for identification, Radio Frequency Identification (RFID) has enabled applications, including inventory management by asset tracking. Later, the Wireless Identification and Sensing Platform (WISP) added digital sensor functionality \cite{sampleDesignRFIDBasedBatteryFree2008}. More recently, tags can passively harvest radio energy while also supporting active wireless transmission \cite{Wiliot2024}.
These advances in tag passive energy harvesting with active transmission stem from using extremely low-power chip design. 
Also benefiting from low-power silicon, sensor tag computation allows encrypting identification and sensor data with an onboard Advanced Encryption Standard (AES) key before actively broadcasting on carrier frequencies that differ from those providing the passive RF energy to the tag.
The tags used for our experiment passively collect wireless energy from 918~MHz and actively transmit on Bluetooth Low Energy (BLE) advertisement channels (CH37, CH38, and CH39) at the 2.4~GHz band. Within the BLE specification, these channels are used exclusively for ad-hoc pairless simplex transmission and are placed between WiFi channel bands. Due to on-tag encryption, data is available through this advertisement broadcast to anyone listening but requires decryption from an online service.

\begin{figure}[ht] 
        \centering
        \includegraphics[width=0.9\linewidth]{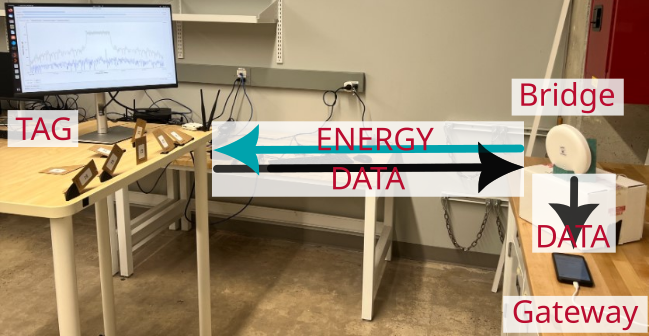} 
        \caption{Gives a lab setup demonstrating how the tags (left) are energized and transmit data to the bridge (right top). The bridge then relays the data to a gateway (cellphone, bottom right) with internet connectivity to decrypt the data via an online service. } 
        \label{fig:lab_setup}
    \end{figure}

Due to low-power design, the active transmission power of the tags is very weak and must be supported by a bridge device with a high-gain directional receiver to detect the tags from a maximum range of 10~m. The bridge receiver relays the BLE packet advertisement data to a gateway device with internet access as a connection-less mesh \cite{lacavaSecuringBluetoothLow2022}. This process is shown in Figure \ref{fig:lab_setup}. The gateway enables pseudo-real-time decryption via an internet service, which retrieves the decoded packet data. Many packet data types are supported and are retrieved from the decryption service: gateway and bridge connection statuses, gateway location, the bridge that detected the tag, whether a tag is active, and the tag’s temperature. This application uses only the tag activity status and temperature data properties; both of these used packet types include the tag's unique ID.

\subsection{AERPAW} \label{sec:AERPAW} 
The Aerial Experimentation and Research Platform for Advanced Wireless (AERPAW) is part of the NSF’s Platforms for Advanced Wireless Research (PAWR) \cite{marojevicAdvancedWirelessUnmanned2020}. It provides wireless researchers access to autonomous aerial drones and fixed towers with various wireless research payloads and Software Defined Radio (SDR) equipment. In addition to the wireless testbed, the platform includes a Digital Twin (DT) emulation environment. This DT environment allows users to safely test and develop autonomous programs on emulated drone hardware and simulated wireless equipment. The DT can functionally emulate both the vehicle and an SDR wireless channel based on the drone's current position. Once a user has developed a validated drone control program within the DT, AERPAW transfers the same containerized algorithm onto physical equipment for trial upon the physical testbed. This DT to testbed development pipeline allows for flexible but repeatable deployment.
The AERPAW platform is also open hardware, and researchers can develop custom wireless device payloads. These non-canonical experiments can use parts of the AERPAW platform to facilitate the rapid testing and development of new wireless technologies, dramatically accelerating the testing of novel wireless communication experiments without redeveloping complex flight systems.

\section{Experimental Methodology} \label{sec:method} 

\begin{figure}[!tp] 
        \centering
        \includegraphics[width=0.9\linewidth]{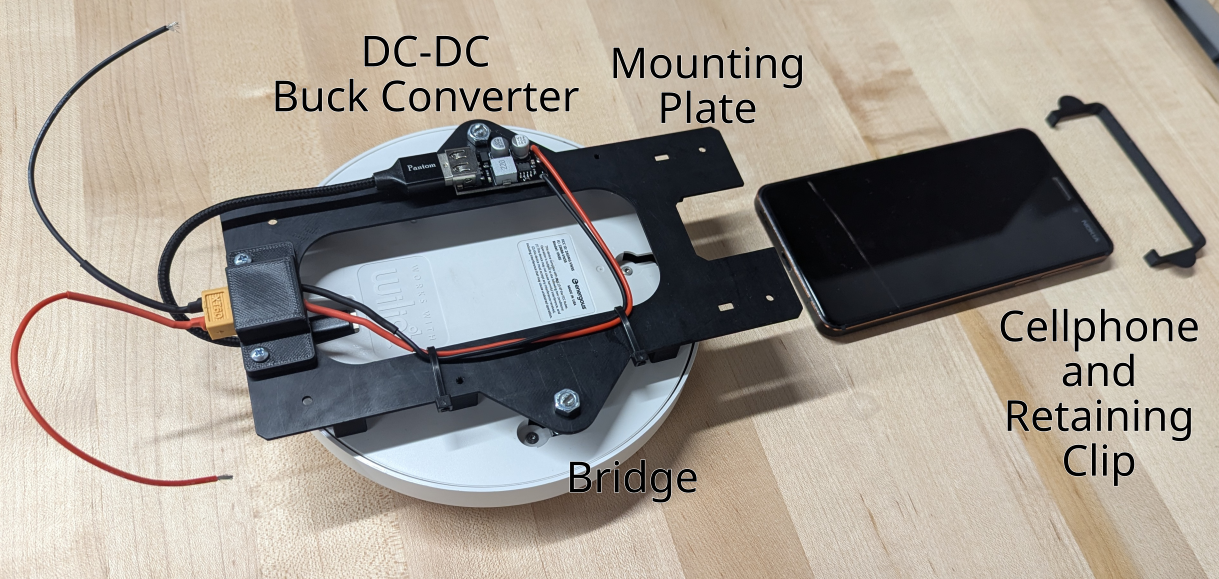} 
        \caption{Shows the experimental wireless platform assembly. The white circular object to the left is the tag bridge with an attached mounting plate and buck converter for power. To the right is the cell phone that slides between the bridge and mounting plate and is held in place with a retaining clip.} 
        \label{fig:mounting_plate}
    \end{figure}

To examine the feasibility of using wireless tags for remote sensor monitoring with a passively powered, actively transmitting tag, a payload platform was constructed to fit on a SAM. This payload consists of four major components: a Nokia 6.1 cellphone, the tag bridge device, a DC-DC buck converter to provide 5V power to the bridge, and a 3D printed platform for mounting the components onto the drone. The payload assembly is shown in Figure \ref{fig:mounting_plate}.  The individual components were placed into Computer-Aided Design (CAD), so a platform connecting the phone, bridge, and converter could be created specifically for 3D printing. The bridge device is rigidly affixed via the payload platform to the drone. However, the phone, which required user access before the experiment was designed to be easily removable. The phone is then retained within the platform with a flexure clip.

    \begin{figure*} 
        \centering
        \includegraphics[width=0.8\linewidth]{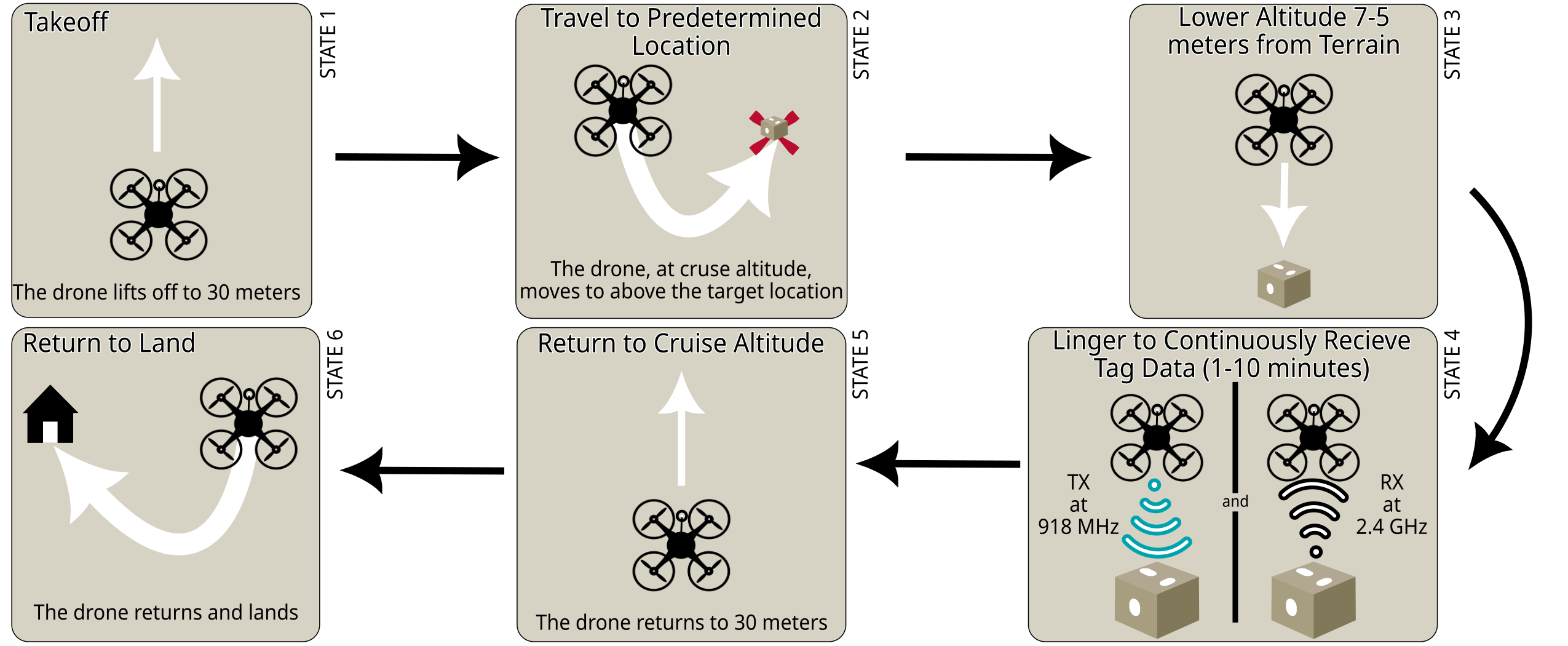} 
        \caption{Shows the state machine flight diagram for the programmed autonomous data collection.} 
        \label{fig:flight_diagram}
    \end{figure*}

With the payload designed, printed, and constructed, a suitable backend infrastructure was created to facilitate collecting and testing the incoming tag data. The payload collects and relays data using local cellular coverage to a cloud service for real-time decryption. The cloud service is directed to send the decrypted data to a Message Queuing Telemetry Transport (MQTT) broker. Three microservice Docker containers are then run offsite, allowing for the decrypted tag data collection, storage, and analysis by dashboard visualization. The tag data is first retrieved from the online service via a custom Python script running within a container that connects to the MQTT broker and an ImpulseDB container. This Python container collects the decrypted tag data, parses it, and then sends it to the database. A Graphana container is then used to query for recent data and provide data visualization dashboards. All three microservices are then orchestrated on a host computer within the lab via Docker Compose. A simple connection to the MQTT broker is used for testing conducted outside the lab to validate if tag activity status and temperature data are captured.

To guide the drone, a custom flight plan was created within the AERPAW emulation environment, leveraging example Python code for drone movement as a state machine. This flight behavior is shown in Figure \ref{fig:flight_diagram}. The autonomous script reads variable experiment mission parameters stored within a Yet Another Markup Language (YAML) file. Within the YAML are a target latitude, longitude, and altitude. The drone altitude must be low enough for the bridge to energize the tags and receive data, but because the drone is commanded to such a low altitude, the script checks the ground height at that location in addition to traditional AERPAW testbed flight boundary checks. This ground height check capability was enabled by parsing part of a GeoTiff SRTM+ dataset to retrieve elevation data of the Lake Wheeler testbed flight zone provided by OpenTopography \cite{nsfopentopographyfacilityGlobalBathymetryTopography2019}. The data is held within the drone and checked before takeoff. The AERPAW DT was then used to debug and validate the autonomous script behavior using QGroundControl, an open-source flight control program.

The script commands the drone to take off to a cruise altitude and move towards the target location. Once it arrives at a cruise altitude above the tags, the drone can safely lower directly to a target altitude and linger until sufficient data is collected. This height is set within the mission parameters YAML and, with adjustment, set to hover safely within communication range above the tags on the ground. The drone can proceed to the next waypoint or land upon reaching a specific linger time or estimated battery state of charge when within the linger state. 

\begin{figure*}
        \centering
        \includegraphics[width=0.9\linewidth]{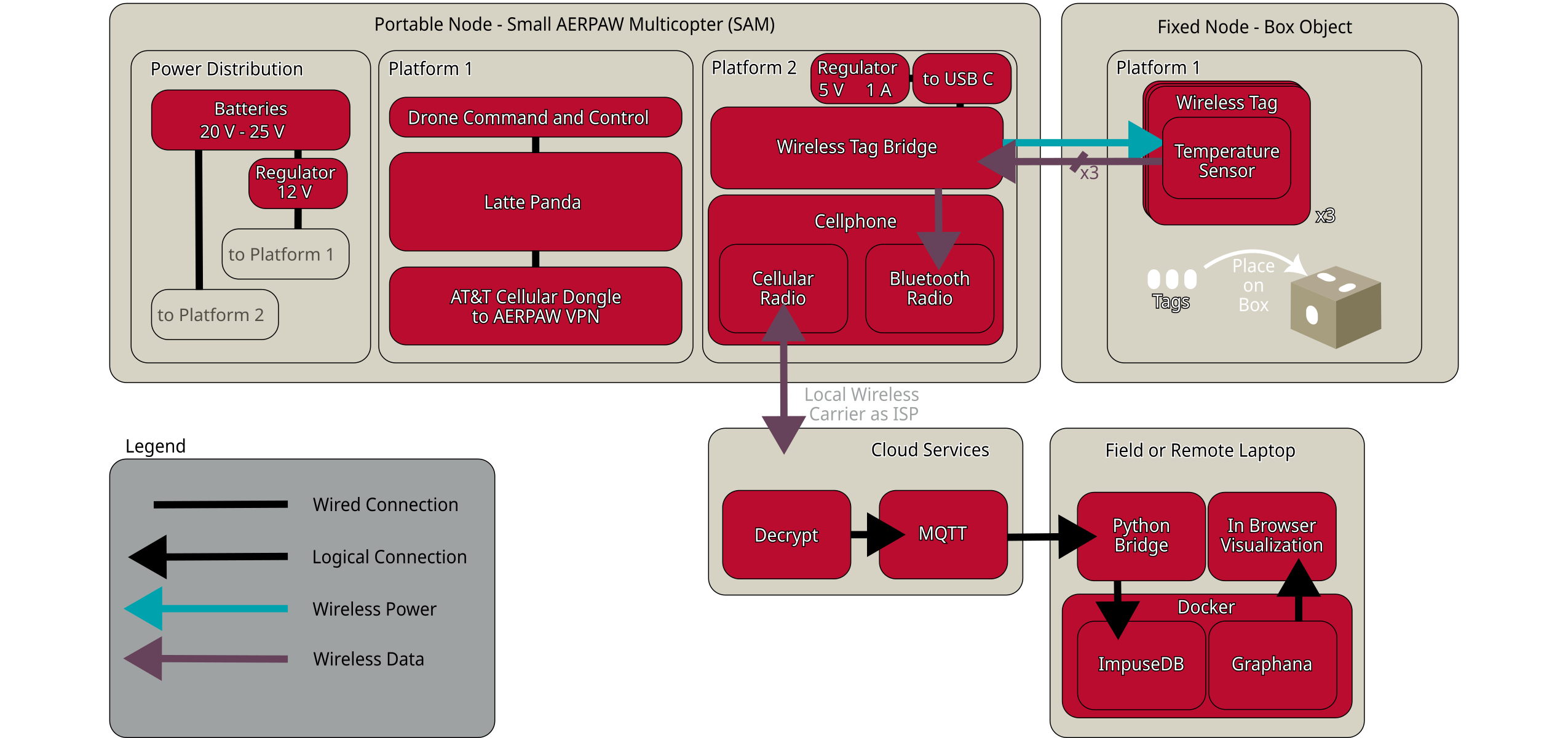} 
        \caption{Shows the block diagram of the small drone. Platform 1 is provided by AERPAW. Platform 2 is the custom payload node. The custom node allows the drone to energize and collect encrypted data packets from the tags, and relay them in real-time to a decryption service. The decrypted data is then stored within a remote database.} 
        \label{fig:system_diagram}
    \end{figure*}

Procedure documentation was produced to assist the AERPAW team in correctly attaching the payload to the aerial drone and bringing the system online before the mission. The payload subassembly was shipped to the AERPAW team for final drone assembly and testbed trials. The trials were planned to proceed in three stages. First, the drone would be powered, and a stationary ground test would be performed. Second, the drone would be powered and manually flown over the target location. After completing the manual flight, it would be given the waypoint latitude and longitude (recorded during the manual flight) and autonomously navigate to the target location. In each experiment flight trial, the drone is flown or commanded to fly directly over three tags taped to a cardboard box placed on the ground with the tags affixed in multiple orientations. The drone's bridge wirelessly energizes the tags and relays the encrypted tag sensor and ID data. The cellphone provides gateway services to the bridge and local cellular connectivity for streaming data to the cloud service in real-time. Data received by MQTT shows tag activity or temperature to the field operators. The system block diagram is shown in Figure \ref{fig:system_diagram}.

\section{Results and Analysis} \label{sec:results} 

Following the 3D printing and construction of the payload, the tag data collection was tested overnight within the lab without issues and shipped to the AERPAW team for installation onto the SAM. Before the flight, following the procedure, the cellphone gateway was activated and placed within the platform. A ground-based subsystem test was conducted. The drone’s systems remained unpowered while the payload was under test. Data was collected indicating good connectivity through the attached payload subsystem. The drone was then manually flown over the box; however, no data was collected. Figure \ref{fig:man_flight} shows the drone being manually flown over the target location.

The drone was then disarmed and held over the box in the field. This test showed data being collected. The cell phone was removed from the payload and strapped to the drone's top to improve connectivity during flight. The drone was again manually flown over the top of the box for several minutes. A single active tag data packet was received during this second manual flight, but no temperature data was successfully received. Because both manual flights were unable to collect temperature data, an autonomous flight was forgone to troubleshoot the data connectivity issue.


To further isolate the cause of the data disruption, an improvised test stand was constructed that allowed the drone to be armed while remaining stationary. Arming the drone allows the motors to spin without generating enough thrust for takeoff. Three tests were conducted on the test stand, and an additional preliminary test was conducted with the drone unarmed as a control. In each test, the cell phone was moved to a different location. The control test was performed first with the drone unarmed, above the box, and the phone was in the default position within the payload. Data was successfully collected with this control test indicating no faults within the test stand setup. The first arming test was performed with the phone in the normal payload position, and no data was collected. This replicated similar conditions to that of the first manual flight. The second test strapped the phone above the drone, replicating the second trial flight. Similarly, no data was collected. The phone was then placed on the table beside the test stand, and in this test, data was collected.

The second test, moving the phone to the top of the drone, isolated the phone from any inadvertent shielding or small-signal Electromagnetic Interference (EMI) from the small buck converter and onboard computer. However, this second test located the phone above the batteries, still well within the EMI effects of the drone power distribution. The third test, with the phone further isolated on the table, positively indicated that drone EMI effectively blocked the communication link between the phone and the bridge.

    \begin{figure}[!ht] 
        \centering
        \includegraphics[width=0.8\linewidth]{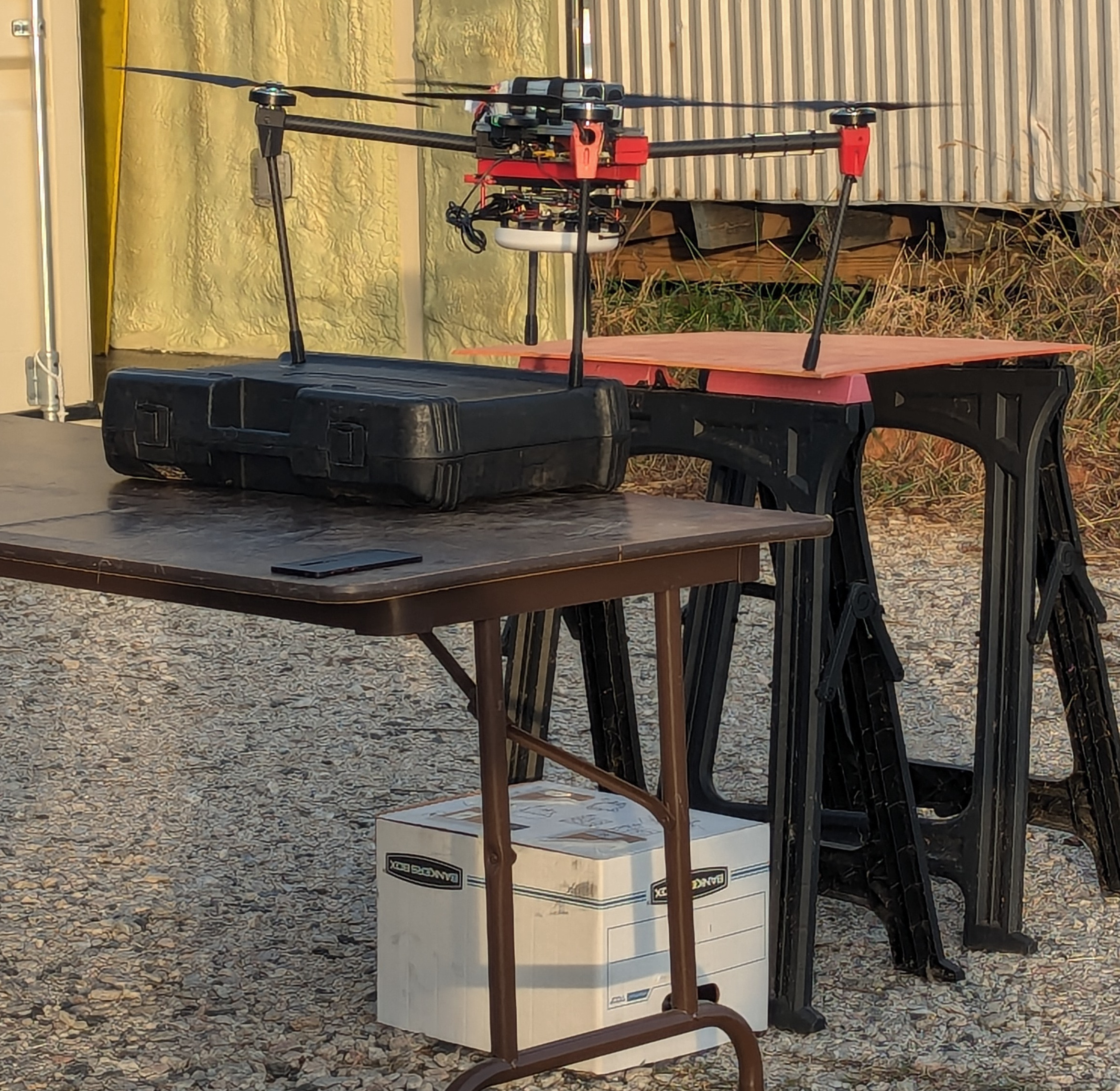} 
        \caption{Shows the drone on an improvised test stand during the third test where the cell phone is placed on the table.} 
        \label{fig:teststand}
    \end{figure}

\section{Conclusion} \label{sec:conclusion} 

The availability of the AERPAW research platform and the support of the AERPAW team greatly facilitated the rapid execution of this experiment. While the full mission completion did not meet the expected results of an autonomous drone flying to collect tag data, it still provided valuable insights from the performed experiments. Specifically, the third test stand experiment indicated a communication breakdown in the cellphone link highly correlated with proximity to drone motor activity. The second manual flight, with the phone on top of the drone, successfully received a tag activity data packet. This result demonstrated the ability of the system to still operate in extreme interference by collecting a tag's ID data, albeit with undesired lengths of linger time while under manual control. A replication of the two manual flight conditions with a test stand control further indicated that the system interference was linked to the drone’s motor operation.

\section{Further Directions} \label{sec:further} 
While the experiment results show that more research is needed to bring this technology to fruition within the challenging field of aerial drones and agriculture, the idea is still promising. Expanding the scope of passively powered, actively transmitting tag research would allow for a more open hardware tag platform, enabling a more heterogeneous selection of sensor modalities and more robust diagnostics for research and testing. Open tag hardware would also allow for condensing the device count in the payload subassembly into a single receiver device. This would lead to an overall system weight reduction, a critical design criterion for aerial drones. A device unification would also simplify the number of wireless data links, reducing potential vector sources of wireless interference and enhancing system reliability. Additionally, because the service used for decrypting the data from the tags requires a timely relay of the encrypted data, the encrypted packets cannot be stored to be decoded later. This limitation would need to be addressed for remote applications with limited or unavailable internet access.



\section*{Acknowledgment} 
The work is supported by NSF CNS-2236449 and its supplemental fund. The authors also express their gratitude to Drs. Ismail Guvenc, Mihail Sichitiu, Ozgur Ozdemir, and the entire AERPAW team for their invaluable contributions to this experiment.  

\bibliographystyle{ieeetr}
\bibliography{cite.bib}






\end{document}